\documentclass[preprint,superscriptaddress,preprintnumbers,showpacs]{elsart}
\usepackage{amssymb}
\usepackage{graphicx}
\usepackage{dcolumn}
\usepackage{bm}

\journal{Physics Letter B}

\begin{document}

\begin{frontmatter}

\title{Unitarity boomerangs of quark and lepton mixing matrices}

\author[1]{Shi-Wen Li},
\author[1,2]{Bo-Qiang Ma\corauthref{*}} \corauth[*]{Corresponding author.}\ead{mabq@phy.pku.edu.cn}
\address[1]{School of Physics and State Key Laboratory of Nuclear Physics and
Technology, Peking University, Beijing 100871, China}
\address[2]{Center for High Energy
Physics, Peking University, Beijing 100871, China}


\begin{abstract}
The most popular way to present mixing matrices of quarks (CKM)
and leptons (PMNS) is the parametrization with three mixing angles
and one CP-violating phase. There are two major options in this kind
of parametrizations, one is the original Kobayashi-Maskawa (KM)
matrix, and the other is the Chau-Keung (CK) matrix. In a new
proposal by Frampton and He, a unitarity boomerang is introduced to
combine two unitarity triangles, and this new presentation displays
all four independent parameters of the KM parametrization in the
quark sector simultaneously. In this paper, we study the relations
between KM and CK parametrizations, and also consider the
quark-lepton complementarity (QLC) in the KM parametrization. The
unitarity boomerang is discussed in the situation of the CK parametrization
for comparison with that in the KM parametrization in the quark sector.
Then we extend the idea of unitarity boomerang to the
lepton sector, and check the corresponding unitarity boomerangs in
the two cases of parametrizations.
\end{abstract}


\end{frontmatter}

\newpage
\section{Introduction}

Mixing of different generations of fermions is one of the most
interesting issues in particle physics. To understand the mixing
patterns and properties, the mixing matrix was introduced for
phenomenological and theoretical studies. In the quark sector, the
mixing matrix is described by the Cabibbo\cite{cabibbo}-Kobayashi-Maskawa\cite{km}(CKM) matrix
$V_{\rm CKM}$, and in the lepton sector, it is described by the
Pontecorvo\cite{pontecorvo}-Maki-Nakawaga-Sakata\cite{mns} (PMNS)
matrix $U_{\rm PMNS}$,
\begin{eqnarray}
V_{\rm CKM}=\left(
  \begin{array}{ccc}
    V_{ud} & V_{us} & V_{ub} \\
    V_{cd} & V_{cs} & V_{cb} \\
    V_{td} & V_{ts} & V_{tb} \\
  \end{array}
\right), \quad
U_{\rm PMNS}=\left(
  \begin{array}{ccc}
    U_{e1}    & U_{e2}    & U_{e3}    \\
    U_{\mu1}  & U_{\mu2}  & U_{\mu3}  \\
    U_{\tau1} & U_{\tau2} & U_{\tau3} \\
  \end{array}
\right).\nonumber
\end{eqnarray}

The original Kobayashi-Maskawa (KM)~\cite{km} matrix was introduced
in 1973 to accommodate CP violation in the Standard Model, by an
extension of the Cabibbo's idea~\cite{cabibbo} of quark mixing from
two generations to three generations with six quark flavors,
\begin{eqnarray}
V_{\rm KM}&=&\left(
  \begin{array}{ccc}
    1 & 0   & 0    \\
    0 & c_2 & -s_2 \\
    0 & s_2 & c_2  \\
  \end{array}
\right)\left(
  \begin{array}{ccc}
    c_1 & -s_1 & 0 \\
    s_1 & c_1  & 0 \\
    0   & 0    & e^{i\delta_{\rm KM}} \\
  \end{array}
\right)\left(
  \begin{array}{ccc}
    1 & 0   & 0    \\
    0 & c_3 & s_3  \\
    0 & s_3 & -c_3 \\
  \end{array}
\right)\nonumber\\
&=&\left(
\begin{array}{ccc}
c_1     & -s_1c_3                     & -s_1s_3           \\
s_1c_2  & c_1c_2c_3-s_2s_3e^{i\delta_{\rm KM}} & c_1c_2s_3+s_2c_3e^{i\delta_{\rm KM}} \\
s_1s_2  & c_1s_2c_3+c_2s_3e^{i\delta_{\rm KM}} & c_1s_2s_3-c_2c_3e^{i\delta_{\rm KM}}
\end{array}
\right).\label{KM}
\end{eqnarray}
Here $s_i=\sin\theta_i$, $c_i=\cos\theta_i$ $(i=1,2,3)$, and $\theta_1$, $\theta_2$, $\theta_3$ are Euler angles, $\delta_{\rm KM}$ is the CP-violating phase in the KM parametrization.

In 1984, Chau and Keung (CK)~\cite{ck} introduced a different
parametrization, which has been advocated by the Particle Data Group
since 1996~\cite{1996,pdg},
\begin{eqnarray}
V_{\rm CK}&=&\left(
\begin{array}{ccc}
1 & 0 & 0           \\
0 & c_{23} & s_{23} \\
0 & -s_{23} & c_{23}\\
\end{array}
\right)\left(
\begin{array}{ccc}
c_{13} & 0 & s_{13}e^{-i\delta_{\rm CK}}           \\
0 & 1 & 0 \\
-s_{13}e^{i\delta_{\rm CK}} & 0 & c_{13}\\
\end{array}
\right)\left(
\begin{array}{ccc}
c_{12}& s_{12} & 0           \\
-s_{12} & c_{12} & 0 \\
0 & 0 & 1\\
\end{array}
\right)\nonumber\\
&=&\left(
\begin{array}{ccc}
c_{12}c_{13} & s_{12}c_{13} & s_{13}e^{-i\delta_{\rm CK}}           \\
-s_{12}c_{23}-c_{12}s_{23}s_{13}e^{i\delta_{\rm CK}} &
c_{12}c_{23}-s_{12}s_{23}s_{13}e^{i\delta_{\rm CK}}  & s_{23}c_{13} \\
s_{12}s_{23}-c_{12}c_{23}s_{13}e^{i\delta_{\rm CK}}  &
-c_{12}s_{23}-s_{12}c_{23}s_{13}e^{i\delta_{\rm CK}} & c_{23}c_{13}
\end{array}
\right),\label{CK}
\end{eqnarray}
where $s_{ij}=\sin\theta_{ij}$, $c_{ij}=\cos\theta_{ij}$ $(i,j=1,2,3)$,
$\theta_{12}$, $\theta_{23}$, $\theta_{13}$ are the rotation angles,
and $\delta_{\rm CK}$ is the CP-violating phase in the CK
parametrization.

Although Eq.~(\ref{KM}) and Eq.~(\ref{CK}) are drawn in the quark
sector, the same form of parametrizations can be also used in the
lepton sector since both CKM and PMNS matrices are unitary
matrices for describing the mixing of fermions. However, if
neutrinos are of Majorana type, there should be an additional
diagonal matrix with two Majorana phases $P={\rm
diag}(e^{i\alpha_1/2},e^{i\alpha_2/2},1)$ multiplied to
Eq.~(\ref{KM}) and Eq.~(\ref{CK}). In this paper, we consider the
neutrinos as Dirac neutrinos, and the presentation of formalisms for
Majorana neutrinos can be derived straightforwardly by including the
additional phases. In the following, we use the superscripts $Q$ and
$L$ to denote the parameters in the quark sector and the lepton
sector respectively if necessary.

The experimental data of the moduli of the matrix elements are very
important to understand the mixing matrices since they constitute
most reliable information of the mixing patterns and properties. For
quark mixing, the ranges of magnitude of the CKM matrix elements
have been very well determined with~\cite{pdg}
\begin{eqnarray} \left(
  \begin{array}{ccc}
    0.97419\pm0.00022             & 0.2257\pm0.0010    & 0.00359\pm0.00016               \\
    0.2256\pm0.0010               & 0.97334\pm0.00023  & 0.0415^{+0.0010}_{-0.0011}      \\
    0.00874^{+0.00026}_{-0.00037} & 0.0407\pm0.0010    & 0.999133^{+0.000044}_{-0.000043}
  \end{array} \right)\;.\label{ckmdata}
\end{eqnarray}
For lepton mixing, the ranges for the PMNS matrix elements have been
also constrained by (at $3\sigma$ level)~\cite{gonzalez}
\begin{eqnarray} \left(
  \begin{array}{ccc}
    0.77 - 0.86      & 0.50 - 0.63      & 0.00 - 0.22      \\
    0.22 - 0.56      & 0.44 - 0.73      & 0.57 - 0.80      \\
    0.21 - 0.55      & 0.40 - 0.71      & 0.59 - 0.82
  \end{array} \right)\;.\label{pmnsdata}
\end{eqnarray}

Now we have the moduli of the elements of the two mixing matrices,
but it is not easy to study these elements from experimental data
directly. For convenience, it is common to adopt an approximation as
the basis matrix to the lowest order. In the quark sector, a better
choice is the unit matrix and/or the matrix suggested in
Ref.~\cite{tripara}
\begin{eqnarray}
V_0=\left(
              \begin{array}{ccc}
                1 & 0 & 0 \\
                0 & 1 & 0 \\
                0 & 0 & 1 \\
              \end{array}
            \right),\quad
    V'_0=\left(
        \begin{array}{ccc}
            \frac{\sqrt{2}+1}{\sqrt{6}} & \frac{\sqrt{2}-1}{\sqrt{6}} & 0 \\
           -\frac{\sqrt{2}-1}{\sqrt{6}} & \frac{\sqrt{2}+1}{\sqrt{6}} & 0 \\
            0                           & 0                           & 1
        \end{array}
        \right).\label{relatri}
\end{eqnarray}
The unit matrix is very simple while the later one is more close to experimental data.

In the lepton sector, it has been common to choose the bimaximal
matrix~\cite{bi} and/or the tri-bimaximal matrix~\cite{tri} as the
basis matrices
\begin{eqnarray}
U_{\rm{bi}}=\left(
              \begin{array}{ccc}
                1/\sqrt{2} & 1/\sqrt{2} & 0 \\
                -1/2       & 1/2        & 1/\sqrt{2} \\
                1/2        & -1/2       & 1/\sqrt{2} \\
              \end{array}
            \right),\quad
U_{\rm{tri}}=\left(
              \begin{array}{ccc}
              2/\sqrt{6}  & 1/\sqrt{3}  & 0          \\
              -1/\sqrt{6} & 1/\sqrt{3}  & 1/\sqrt{2} \\
              1/\sqrt{6}  & -1/\sqrt{3} & 1/\sqrt{2}
              \end{array}
            \right).
\end{eqnarray}
Although the former one is not favored by present experimental data
as the later one, it looks more symmetric with also possible
connection with the unit basis in quark mixing~\cite{qlcparabi}. The
tribimaximal basis is very close to experimental data and can serve
as a good approximation for lepton mixing.

Despite the fact that mixing of quarks and leptons could be treated separately,
it would be interesting to find an internal expression which
connects the two sectors. To this end, the quark-lepton
complementarity (QLC)~\cite{qlc} provides a very useful relation
between quark and lepton mixing angles and leads to a unified
treatment of mixing in quark and lepton sectors. With the QLC, the
unified parametrization of quark and lepton mixing matrices has
been also discussed~\cite{tripara,qlcparabi,qlcparatri}. The QLC reads as the
following equations
\begin{eqnarray}
\theta_{12}^Q+\theta_{12}^L=\frac{\pi}{4},
\quad\theta_{23}^Q+\theta_{23}^L=\frac{\pi}{4},
\quad\theta_{31}^Q\sim\theta_{31}^L\sim 0,\label{qlc}
\end{eqnarray}
where the parameters $\theta^{Q,L}_{12}$, $\theta^{Q,L}_{23}$,
$\theta^{Q,L}_{13}$ refer to the rotation angles in the CK
parametrization. Under the QLC, it is interesting to find that the
unit matrix $V_0$ in the quark sector corresponds to the bimaximal
matrix $U_{\rm bi}$ in the lepton sector~\cite{qlcparabi}, which enlightened
us to search for the corresponding matrix in the quark sector to the tri-bimaximal
matrix in the lepton sector. That is the matrix $V'_0$ in Eq.~(\ref{relatri}), which
is drawn with the combination of the QLC and the tri-bimaximal matrix
$U_{\rm tri}$~\cite{tripara}. In other words, the
basis matrix $V'_0$ in the quark sector
corresponds to the tri-bimaximal matrix $U_{\rm tri}$ under the QLC in the lepton
sector~\cite{qlcparatri,He:2008td}.

The unitarity of the mixing matrix imposes six vanishing
combinations which can be represented in a complex plane as
triangles. They are well known as the unitarity triangles, which
play an important role in understanding the mixing matrices. There
are three inner angles and three sides in one unitarity triangle,
however, only three of them are independent. To give all four
parameters in the mixing matrix, one must take another unitarity
triangle into account which leads to the idea of unitarity boomerang
introduced by Frampton and He~\cite{fh}. They also suggested that
the four independent parameters of the unitarity boomerang in the KM
parametrization are convenient for phenomenological studies. For a
systematic study of this issue, we present in Sec.~II the relations
between KM and CK parametrizations, and discuss the quark-lepton
complementarity (QLC) under the KM parametrization. In Sec.~III, we study
unitarity boomerangs in the quark sector for both KM and CK
parametrizations. The shapes of boomerangs in the two cases are
compared, and our analysis supports the proposal by Frampton and He.
In Sec.~IV, we also extend the idea of unitarity boomerang to the
lepton sector, and study boomerangs in both parametrizations.
Finally we give a summary in Sec.~V.

\section{Relations between KM and CK parametrizations and the quark-lepton complementarity in the KM parametrization}

Now let us check the relations between KM and CK
parametrizations. Following the steps in Ref.~\cite{ck}, we redefine
the fields of the $c$ quark, $t$ quark, $s$ quark and $b$ quark by
\begin{eqnarray}
c\rightarrow ce^{i(\pi+\sigma_c)},\quad t\rightarrow te^{i(\pi+\sigma_t)},\quad s\rightarrow se^{i\pi},\quad b\rightarrow be^{i(\pi+\delta_{\rm CK})}\nonumber
\end{eqnarray}
so that
\begin{eqnarray}
V_{\rm KM}&\rightarrow&V'_{\rm KM}=\left(
  \begin{array}{ccc}
    1 & 0              & 0    \\
    0 & -e^{i\sigma_c} & 0    \\
    0 & 0              & -e^{i\sigma_t}  \\
  \end{array}
\right)V_{\rm KM}\left(
  \begin{array}{ccc}
    1 & 0   & 0    \\
    0 & -1   & 0    \\
    0 & 0   & -e^{-i\delta_{\rm CK}} \\
  \end{array}
\right)\nonumber\\
&=&\left(
\begin{array}{ccc}
c_1     & s_1c_3                     & s_1s_3e^{-i\delta_{\rm CK}}           \\
-s_1c_2e^{i\sigma_c}  & c_1c_2c_3e^{i\sigma_c}-s_2s_3e^{i(\sigma_c+\delta_{\rm KM})} & |c_1c_2s_3+s_2c_3e^{i\delta_{\rm KM}}| \\
-s_1s_2e^{i\sigma_t}  & c_1s_2c_3e^{i\sigma_t}+c_2s_3e^{i(\sigma_t+\delta_{\rm KM)}} & |c_1s_2s_3-c_2c_3e^{i\delta_{\rm KM}}|
\end{array}
\right).\label{KMtans}
\end{eqnarray}
where
\begin{eqnarray}
e^{i\sigma_c}=\frac{c_1c_2s_3+s_2c_3e^{-i\delta_{\rm KM}}}{|c_1c_2s_3+s_2c_3e^{i\delta_{\rm KM}}|}e^{i\delta_{\rm CK}},\quad
e^{i\sigma_t}=\frac{c_1s_2s_3-c_2c_3e^{-i\delta_{\rm KM}}}{|c_1s_2s_3-c_2c_3e^{i\delta_{\rm KM}}|}e^{i\delta_{\rm CK}}.
\end{eqnarray}

To transform the KM parametrization into the CK parametrization, the relations of the parameters between the two parametrizations are derived~\cite{fritzsch}:
\begin{eqnarray}
&&s_{12}=s_1c_3(1-s_1^2s_3^2)^{-{1\over2}},\nonumber\\
&&s_{23}=(s_2^2c_3^2+c_1^2c_2^2s_3^2+2c_1c_2c_3s_2s_3\cos\delta_{\rm KM})^{1\over2}(1-s_1^2s_3^2)^{-{1\over2}},\nonumber\\
&&s_{13}=s_1s_3,\nonumber\\
&&s_{23}c_{23}\sin\delta_{\rm CK}=s_2c_2\sin\delta_{\rm KM}.\label{relation}
\end{eqnarray}
It is noted that we use the scripts ``$i$" for quantities in the KM
parametrization and ``$ij$" for those in the CK parametrization.

With the QLC, i.e. the relations between quark and lepton mixing angles in the mode of the CK parametrization, it is interesting to consider the relations of the Euler angles between quark and lepton sectors in the mode of KM parametrization. Fortunately, we have extracted the relations between the corresponding parameters in KM and CK parametrizations in Eq.~(\ref{relation}). Then we can see if there is any similar relations between quark and lepton mixing angles in the KM parametrization.

First, it is straightforward to consider the behavior of the mixing angles under the approximation of the basis matrices.
One finds clearly that the corresponding mixing angles in the two parametrizations
are the same when one only considers the basis matrices, i.e. taking the unit matrix $V_0$ as the basis matrix in the quark sector and
the bimaximal matrix $U_{\rm bi}$ as the basis matrix in the lepton sector, one has
\begin{eqnarray}
&&\theta^Q_1=\theta^Q_{12}=0,\quad\theta^L_1=\theta^L_{12}={\pi\over4},\nonumber\\
&&\theta^Q_2=\theta^Q_{23}=0,\quad\theta^L_2=\theta^L_{23}={\pi\over4},\nonumber\\
&&\theta^Q_3=\theta^Q_{13}=0,\quad\theta^L_3=\theta^L_{13}=0.\nonumber
\end{eqnarray}
Taking $V'_0$ in Eq.~(\ref{relatri}) as the basis matrix in the quark sector and
the the tri-bimaximal matrix $U_{\rm tri}$ as the basis matrix in the lepton sector, one obtains
\begin{eqnarray}
&&\theta^Q_1=\theta^Q_{12}=\arcsin\frac{\sqrt{2}-1}{\sqrt{6}},\quad\theta^L_1=\theta^L_{12}=\arcsin{1\over\sqrt{3}},\nonumber\\
&&\theta^Q_2=\theta^Q_{23}=0,\quad\theta^L_2=\theta^L_{23}={\pi\over4},\nonumber\\
&&\theta^Q_3=\theta^Q_{13}=0,\quad\theta^L_3=\theta^L_{13}=0.\nonumber
\end{eqnarray}

It is consistent with the relations in Eq.~(\ref{relation}). Anyway, the
result means that the QLC relations revealed~\cite{qlc} in the CK
parametrization as expressed in Eq.~(\ref{qlc}) are also satisfied
to the lowest order in the KM parametrization.

Now let us turn to the mixing angles in the realistic mixing matrices. According to Eq.~(\ref{ckmdata}), we find that $s^Q_1\sim \mathcal {O}(10^{-1})$, $s^Q_2\sim \mathcal {O}(10^{-2})$ and $s^Q_3\sim \mathcal {O}(10^{-2})$. Then we can simplify Eq.~(\ref{relation}) as
\begin{eqnarray}
&&s^Q_{12}=s^Q_1+\mathcal {O}(10^{-4}),\nonumber\\
&&s^Q_{23}=\left((s^Q_2)^2+(s^Q_3)^2+2s^Q_2s^Q_3\cos\delta^Q_{\rm KM}\right)^{1/2}+\mathcal {O}(10^{-4}),\nonumber\\
&&s^Q_{13}=s^Q_1s^Q_3,\nonumber\\
&&s^Q_{23}\sin\delta^Q_{\rm CK}=s^Q_2\sin\delta^Q_{\rm KM}+\mathcal {O}(10^{-4}),\label{qsim}
\end{eqnarray}
which has been already obtained by Chau and Keung in Ref.~\cite{ck}.

As to the quark sector, we find that the approximation is not so
simple according to Eq.~(\ref{pmnsdata}) in the lepton sector.
Nevertheless, $|U_{e3}|$ being small implies that $s^L_3$ is a small
parameter. Thus, we could expand the equations in
Eq.~(\ref{relation}) in powers of $s^L_3$ and obtain
\begin{eqnarray}
&&s^L_{12}=s^L_1-{1\over2}s^L_1(c^L_1)^2(s^L_3)^2+\mathcal {O}\left((s^L_3)^4\right),\nonumber\\
&&s^L_{23}=s^L_2+c^L_1c^L_2\cos\delta^L_{\rm KM}s^L_3+\frac{(c^L_1)^2}{2s^L_2}\left((c^L_2)^2-(s^L_2)^2-2(c^L_2)^2\cos^2\delta^L_{\rm KM}\right)(s^L_3)^2+\mathcal {O}\left((s^L_3)^3\right),\nonumber\\
&&s^L_{13}=s^L_1s^L_3,\nonumber\\
&&s^L_{23}c^L_{23}\sin\delta^L_{\rm CK}=s^L_2c^L_2\sin\delta^L_{\rm
KM}.\label{lsim}
\end{eqnarray}

From Eq.~(\ref{pmnsdata}), we know that $s^L_3$ is at most of order
$\mathcal {O}(10^{-1})$, but we do not know whether it is small
enough. Combining Eq.~(\ref{qsim}) with Eq.~(\ref{lsim}), we find
that:
\begin{enumerate}
\item Since $s^L_3$ is small, we always have
\begin{eqnarray}
\theta_3^Q\sim \theta_3^L\sim\theta_{13}^Q\sim \theta_{13}^L\sim 0.\nonumber
\end{eqnarray}

\item When $s^L_3\sim \mathcal {O}(10^{-1})$, we have
\begin{eqnarray}
\theta_1^Q+\theta_1^L=\theta_{12}^Q+\theta_{12}^L+\mathcal {O}(10^{-2})\sim\frac{\pi}{4}+\mathcal {O}(10^{-2}).\nonumber\\
\theta_2^Q+\theta_2^L=\theta_{23}^Q+\theta_{23}^L+\mathcal {O}(10^{-1})\sim\frac{\pi}{4}+\mathcal {O}(10^{-1}).\nonumber
\end{eqnarray}

\item When $s^L_3\sim \mathcal {O}(10^{-2})$ or smaller, we could obtain
\begin{eqnarray}
\theta_1^Q+\theta_1^L=\theta_{12}^Q+\theta_{12}^L+\mathcal {O}(10^{-4})\sim\frac{\pi}{4}+\mathcal {O}(10^{-4}).\nonumber\\
\theta_2^Q+\theta_2^L=\theta_{23}^Q+\theta_{23}^L+\mathcal {O}(10^{-2})\sim\frac{\pi}{4}+\mathcal {O}(10^{-2}).\nonumber
\end{eqnarray}
\end{enumerate}

As discussed above, the QLC is still an appealing relation between
quarks and leptons in the KM parametrization when $s^L_3\sim
\mathcal {O}(10^{-2})$ or smaller. When $s^L_3\sim \mathcal
{O}(10^{-1})$, we find that the QLC in the KM parametrization is not
as good as it does in the CK parametrization. Nevertheless, the QLC
with Euler angles is satisfied to the lowest order in the KM
parametrization. To higher order, the relation may be corrected to
some extent. More precision experimental data are needed to test the
QLC in both CK and KM parametrizations. Knowledge of QLC for
both cases can provide us more complete information concerning the
possible connection and unification between quarks and leptons.

\section{Unitarity boomerangs in quark sector}

We know that the unitarity of the mixing matrix imposes six
unitarity triangles and the areas of them are the same, equal to
half of the Jarlskog invariant $J$~\cite{jarlskog} defined by
\begin{eqnarray}
{\rm Im}(V_{ij}V_{kl}V_{il}^*V_{kj}^*)=J\sum_{m,n}\epsilon_{ikm}\epsilon_{jln},\label{jar}
\end{eqnarray}
which, as an important parameter to the mixing matrix, is
phase-convention independent when measuring CP violation. Therefore,
the inner angles of the unitarity triangles could be related to the
parameter $J$.

The sides of the unitarity triangles are decided by the values of
the matrix elements, so they are important to find the shape of the
unitarity triangles. To understand better, we should know the orders
of magnitude of the matrix elements in the first step of analysis.

The Wolfenstein parametrization~\cite{wolfensteinpara} displays a
good hierarchy among the nine elements of the CKM matrix.
\begin{eqnarray}
V=\left(
  \begin{array}{ccc}
    1-\frac{1}{2}\lambda^2   & \lambda                & A\lambda^3(\rho-i\eta) \\
    -\lambda                 & 1-\frac{1}{2}\lambda^2 & A\lambda^2             \\
    A\lambda^3(1-\rho-i\eta) & -A\lambda^2            & 1                      \\
  \end{array}
\right)+\mathcal{O}(\lambda^4),\label{wolfenstein}
\end{eqnarray}
with $\lambda=0.2257^{+0.0009}_{-0.0010}$ and $A=0.814^{+0.021}_{-0.022}$~\cite{pdg}. It suggests that the CKM matrix could be simply presented as
\begin{eqnarray}
\left(
\begin{array}{ccc}
1          & \lambda   & \lambda^3 \\
\lambda    & 1         & \lambda^2 \\
\lambda^3  & \lambda^2 & 1
\end{array}
\right),\nonumber
\end{eqnarray}
with $\lambda\sim0.2$. It is natural to see that the three sides are
of the same order $\lambda^3$ only in two unitarity triangles arise
from
\begin{eqnarray}
V_{ud}V^*_{ub}+V_{cd}V^*_{cb}+V_{td}V_{tb}^*=0,\quad V_{ud}V^*_{td}+V_{us}V^*_{ts}+V_{ub}V_{tb}^*=0.\label{ckmtwo}
\end{eqnarray}
While in the last four unitarity triangles, one side is
$\mathcal{O}(\lambda^2)$ or $\mathcal{O}(\lambda^4)$ to the other
two sides.

We know that one must take two unitarity triangles to give all four
independent parameters in the mixing matrix. For this purpose,
Frampton and He introduced a new diagram for the quark mixing matrix in
Ref.~\cite{fh}, the unitarity boomerang. For one inner angle of a
unitarity triangle, we can always find a same angle in another
unitarity triangle with the Jarlskog invariant $J$. With the common
angle in overlap for the two triangles, a unitarity boomerang is
then constructed. We have stated that the three sides of the same
order of magnitude only exist in two unitarity triangles arising
from Eq.~(\ref{ckmtwo}). So the unitarity boomerang consisted by the
two unitarity triangles arising from Eq.~(\ref{ckmtwo}) is the most
convenient one, and this is just the choice by Frampton and
He~\cite{fh}. Since the common angle of the two chosen unitarity
triangles could be determined by the CP-violating measurement $J$,
the CP-violating phase could then be constrained.

\begin{figure}[hb]
\begin{center}
\scalebox{0.6}{\includegraphics[100,580][470,770]{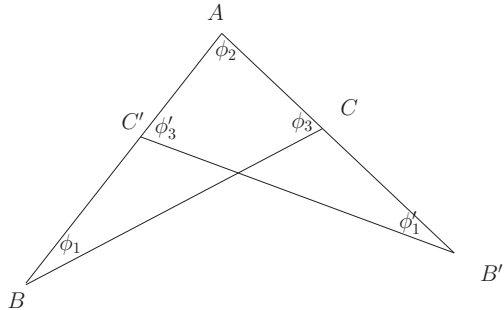}}
\caption{The unitarity boomerang in the quark sector with the common
angle $\phi_2$. The sides are: $AC=|(V_{ud})_{\rm KM}(V^*_{ub})_{\rm
KM}|$, $AC'=|(V_{ub})_{\rm KM}(V^*_{tb})_{\rm KM}|$,
$AB=|(V_{td})_{\rm KM}(V^*_{tb})_{\rm KM}|$, $AB'=|(V_{ud})_{\rm
KM}(V^*_{td})_{\rm KM}|$, $BC=|(V_{cd})_{\rm KM}(V^*_{cb})_{\rm
KM}|$ and $B'C'=|(V_{us})_{\rm KM}(V^*_{ts})_{\rm KM}|$.
\label{fig1}}
\end{center}
\end{figure}

Now we consider the unitarity boomerangs in the KM and CK parametrizations:

{\it Case 1}. The unitarity boomerang in the KM parametrization. As
discussed in Ref.~\cite{fh}, the Jarlskog parameter satisfies
\begin{eqnarray}
J^Q&=&2|(V_{td})_{\rm KM}(V^*_{tb})_{\rm KM}||(V_{ud})_{\rm KM}(V^*_{ub})_{\rm KM}|\sin\phi_2\nonumber\\
&=&2|(V_{ud})_{\rm KM}(V^*_{td})_{\rm KM}||(V_{ub})_{\rm KM}(V^*_{tb})_{\rm KM}|\sin\phi'_2,\nonumber
\end{eqnarray}
with $\phi_2=\phi'_2$ as the common angle there is the diagram of
unitarity boomerang, as illustrated in Fig.~\ref{fig1}.

According to Eq.~(\ref{ckmdata}), we can estimate that $AB\sim AB'$, $AC\sim AC'$, $BC\sim B'C'$, $AB\sim2.5AC$, and $BC\sim2.6AC$.

The CP-violating phase in the KM parametrization is also constrained in Ref.~\cite{fh}
\begin{eqnarray}
\delta^Q_{\rm KM}\approx\pi-\phi_2\approx 90^\circ.\nonumber
\end{eqnarray}

{\it Case 2}. The unitarity boomerang in the CK parametrization. In
this case, to find the constraint of the CP-violating phase
$\delta^Q_{\rm CK}$, we have to choose another unitarity triangle
arising from
\begin{eqnarray}
V_{ud}V^*_{cd}+V_{us}V^*_{cs}+V_{ub}V_{cb}^*=0.\label{ckm3}
\end{eqnarray}

\begin{figure}[hb]
\begin{center}
\scalebox{0.75}{\includegraphics[70,660][540,770]{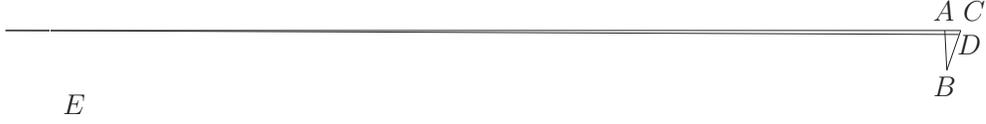}} \caption{The
unitarity boomerang in the quark sector with the common angle $\phi_3$. $\phi_3=\angle ACB=\angle DCE$.
The sides are: $AC=|(V_{ud})_{\rm CK}(V^*_{ub})_{\rm CK}|$,
$AB=|(V_{td})_{\rm CK}(V^*_{tb})_{\rm CK}|$, $BC=|(V_{cd})_{\rm
CK}(V^*_{cb})_{\rm CK}|$, $CD=|(V_{ub})_{\rm CK}(V^*_{cb})_{\rm
CK}|$, $CE=|(V_{ud})_{\rm CK}(V^*_{cd})_{\rm CK}|$ and
$DE=|V_{us})_{\rm CK}(V^*_{cs})_{\rm CK}|$. \label{fig2}}
\end{center}
\end{figure}

We use the Jarlskog parameter expressed as
\begin{eqnarray}
J^Q&=&2|(V_{cd})_{\rm CK}(V^*_{cb})_{\rm CK}||(V_{ud})_{\rm CK}(V^*_{ub})_{\rm CK}|\sin\phi_3\nonumber\\
&=&2|(V_{ud})_{\rm CK}(V^*_{cd})_{\rm CK}||(V_{ub})_{\rm CK}(V^*_{cb})_{\rm CK}|\sin\phi'_3,\nonumber
\end{eqnarray}
so we have the diagram with $\phi_3=\phi'_3$ as the common angle, as
illustrated in Fig.~\ref{fig2}.
\newpage

In Fig.~\ref{fig2} the unitarity triangle $ABC$ is still the same
one in Fig.~\ref{fig1} and we find that the side $CE$ almost coincides with the side $DE$ because of the estimation $CE\sim DE\sim60AC$ and $CD\sim0.043AC$ according to Eq.~(\ref{ckmdata}).

Using experimental values for $|V_{us}|=0.97419\pm 0.00022$, $|V_{ub}| = 0.00359\pm 0.00016$ and $|V_{cb}| = 0.0415^{+0.0010}_{-0.0011}$ in Eq.~(\ref{ckmdata}), one finds that $c^Q_{12}s^Q_{23}s^Q_{13}\ll1$. At a few percent level, one has $(V_{cd})_{\rm CK}=(-s^Q_{12}c^Q_{23}-c^Q_{12}s^Q_{23}s^Q_{13} e^{i\delta^Q_{\rm CK}})\approx -s^Q_{12}c^Q_{23}$. Then
\begin{eqnarray}
\phi_3=\arg\left(-{c^Q_{12}c^Q_{13}s^Q_{13}e^{i\delta^Q_{\rm CK}}\over (-s^Q_{12}c^Q_{23}-c^Q_{12}s^Q_{23}s^Q_{13}e^{i\delta^Q_{\rm CK}})s^Q_{23} c^Q_{13}}\right)\approx
\arg\left({c^Q_{12}c^Q_{13}s^Q_{13}e^{i\delta^Q_{\rm CK}}\over s^Q_{12}c^Q_{23}s^Q_{23}c^Q_{13}}\right)=\delta^Q_{\rm CK}.\nonumber
\end{eqnarray}
The CP-violating phase $\delta^Q_{\rm CK}$ in the CK parametrization is equal to $\phi_3$ to a good approximation. The fact that $\phi_3=(77^{+30}_{-32})^\circ$~\cite{pdg} implies $\delta^Q_{\rm CK}\approx (77^{+30}_{-32})^\circ$.

In Ref.~\cite{fh}, Frampton and He also gave an example to explain
how the unitarity boomerang presents the four independent parameters
in the CKM matrix. They chose three sides $a$, $b$, $c$ of the
unitarity triangle $ABC$ and a side $d$ of the unitarity triangle
$AB'C'$ in Fig.~\ref{fig1} as the four parameters of the CKM matrix
and obtained their expressions with the KM parameters.
\begin{eqnarray}
&&a = |(V_{ud})_{\rm KM}(V^*_{ub})_{\rm KM}|=c^Q_1s^Q_1s^Q_3;\nonumber\\
&&b = |(V_{cd})_{\rm KM}(V^*_{cb})_{\rm KM}|=s^Q_1c^Q_2|c^Q_1c^Q_2s^Q_3+s^Q_2c^Q_3e^{-i\delta^Q_{\rm KM}}|;\nonumber\\
&&c = |(V_{td})_{\rm KM}(V^*_{tb})_{\rm KM}|=s^Q_1s^Q_2|c^Q_1s^Q_2s^Q_3-c^Q_2c^Q_3e^{-i\delta^Q_{\rm KM}}|;\nonumber\\
&&d = |(V_{ud})_{\rm KM}(V^*_{td})_{\rm KM}|=c^Q_1s^Q_1s^Q_2.\label{boomkm}
\end{eqnarray}

Considering the relations in Eq.~(\ref{relation}), we have the expressions with the CK parameters (In fact, we can derive them from the CK parametrization directly).
\begin{eqnarray}
&&a = |(V_{ud})_{\rm CK}(V^*_{ub})_{\rm CK}|=c^Q_{12}c^Q_{23}s^Q_{13};\nonumber\\
&&b = |(V_{cd})_{\rm CK}(V^*_{cb})_{\rm CK}|=s^Q_{23}c^Q_{13}|s^Q_{12}c^Q_{23}+c^Q_{12}s^Q_{23}s^Q_{13}e^{i\delta^Q_{\rm CK}}|;\nonumber\\
&&c = |(V_{td})_{\rm CK}(V^*_{tb})_{\rm CK}|=c^Q_{23}c^Q_{13}|s^Q_{12}s^Q_{23}-c^Q_{12}c^Q_{23}s^Q_{13}e^{i\delta^Q_{\rm CK}}|;\nonumber\\
&&d = |(V_{ud})_{\rm CK}(V^*_{td})_{\rm CK}|=c^Q_{12}c^Q_{13}|s^Q_{12}s^Q_{23}-c^Q_{12}c^Q_{23}s^Q_{13}e^{-i\delta^Q_{\rm CK}}|.\label{boomck}
\end{eqnarray}

In this section, we should realize that the unitarity boomerangs
chosen are not arbitrary. In the quark sector, the unitarity
triangle $ABC$ in Fig.~\ref{fig1} is the most commonly used one.
Since it is one of the only two unitarity triangles in which the
three sides are of the same order among all the six unitarity
triangles. With another unitarity triangle $AB'C'$ in
Fig.~\ref{fig1} in which the three sides are of the same order, we
find it more remarkable to introduce the CP-violating phase
$\delta^L_{\rm KM}$ in the KM parametrization. It is natural to
choose the two special unitarity triangles to construct the
unitarity boomerang, as Frampton and He did in Ref.~\cite{fh}. And
there may be some profound implications with the CP-violating phase
$\delta^L_{\rm KM}$ drawn from the special unitarity boomerang. To
manifest the CP-violating phase $\delta^L_{\rm CK}$ in the CK
parametrization, we have to introduce a third unitarity triangle
$CDE$ in Fig.~\ref{fig2}. Then we find that the shape of the
unitarity boomerang in Fig.~\ref{fig1} looks much normal than that
in Fig.~\ref{fig2}. Thus, the CP-violating phase $\delta^Q_{\rm KM}$
in the KM parametrization is more convenient to be constrained than
$\delta^Q_{\rm CK}$ in the CK parametrization with unitarity
boomerang.

\section{Unitarity boomerangs in lepton sector}

Since both the CKM matrix for quarks and the PMNS matrix for leptons
are unitary, we can extend the analysis of unitarity triangles to
the PMNS matrix in correspondence to those in the quark sector. In
the lepton sector, the hierarchy of the matrix elements is not so
evident as that in the quark sector, however, the tri-bimaximal
matrix characterizes the PMNS matrix pretty well, which means that
the elements in the PMNS matrix are of the same order except
$U_{e3}$. From Eq.~(\ref{pmnsdata}), we only know that $|U_{e3}|$ is
small, $|U_{e3}|\lesssim0.2$, but we do not know whether it is small
enough. We may take $|U_{e3}|\sim0.1$ as an approximation, then the
three sides are nearly of the same order in all six unitarity
triangles in the lepton sector.

Corresponding to the quark sector, we consider two unitarity
triangles arising from
\begin{eqnarray}
U_{e1}U^*_{e3}+U_{\mu1}U^*_{\mu3}+U_{\tau1}U_{\tau3}^*=0,\quad U_{e1}U^*_{\tau1}+U_{e2}U^*_{\tau2}+U_{e3}U_{\tau3}^*=0.\label{pmnstwo}
\end{eqnarray}
The inner angles defined by the two unitarity triangles are
\begin{eqnarray}
\varphi_1&=&\arg \left(-{U_{\mu1}U^*_{\mu3} \over U_{\tau1}U^*_{\tau3}}\right), \nonumber\\
\varphi_2&=&\arg \left(-{U_{\tau1}U^*_{\tau3} \over U_{e1}U^*_{e3}}\right), \nonumber \\
\varphi_3&=&\arg \left(-{U_{e1}U^*_{e3} \over
U_{\mu1}U^*_{\mu3}}\right), \label{UTa2}
\end{eqnarray}
and
\begin{eqnarray}
\varphi'_1 &=& \arg \left( -{U_{e1}U^*_{\tau1} \over U_{e2}U^*_{\tau2}}\right), \nonumber \\
\varphi'_2 &=& \arg \left( -{U_{e3}U^*_{\tau3} \over U_{e1}U^*_{\tau1}}\right), \nonumber \\
\varphi'_3 &=& \arg \left(-{ U_{e2}U^*_{\tau2} \over
U_{e3}U^*_{\tau3}}\right). \label{UTb2}
\end{eqnarray}

Since the experimental data about neutrinos are not so accurate as
those of quarks and we do not know the experimental data about the
inner angles of the unitary triangles of the lepton mixing matrix, it is not easy to find the shape of the unitarity triangles or the unitarity boomerangs. To understand more clearly, we
may take the tri-bimaximal matrix with $|U_{e3}|\lesssim0.2$ as an approximation of the PMNS matrix for example to see what we can learn from the unitarity boomerangs. However, this is only a special case, and the inner angles of the unitarity triangles can not be determined in common cases. In this special case, we obtain
\begin{eqnarray}
\varphi_1\lesssim33^\circ,\quad \varphi_2\sim\varphi_3\gtrsim74^\circ;\quad \varphi'_1\lesssim24^\circ,\quad \varphi'_2\sim\varphi'_3\gtrsim78^\circ.
\end{eqnarray}

Now we attain two isosceles triangles which are built on the special case.
Though the magnitude of the CP-violating phase is still unknown in
the lepton sector, it may be constrained approximately. We still consider the
behavior of the unitarity boomerangs in KM and CK
parametrizations, respectively.

\begin{figure}[hb]
\begin{center}
\scalebox{0.7}{\includegraphics[101,580][473,770]{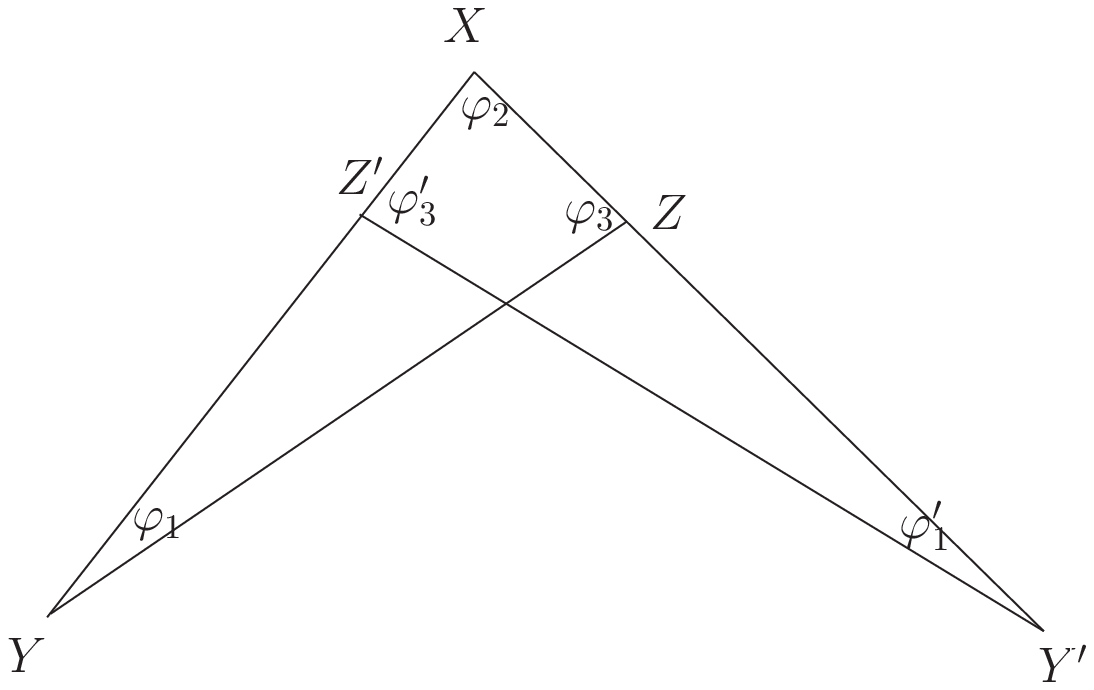}}
\caption{The unitarity boomerang in the lepton sector with the
common angle $\varphi_2$. The sides are: $XZ=|(U_{e1})_{\rm
KM}(U^*_{e3})_{\rm KM}|$, $XY=|(U_{\tau1})_{\rm
KM}(U^*_{\tau3})_{\rm KM}|$, $YZ=|(U_{\mu1})_{\rm
KM}(U^*_{\mu3})_{\rm KM}|$, $XZ'=|(U_{e3})_{\rm
KM}(U^*_{\tau3})_{\rm KM}|$, $XY'=|(U_{e1})_{\rm
KM}(U^*_{\tau1})_{\rm KM}|$ and $Y'Z'=|(U_{e2})_{\rm
KM}(U^*_{\tau2})_{\rm KM}|$. \label{fig3}}
\end{center}
\end{figure}

{\it Case 1'}. The unitarity boomerang in the KM parametrization. The Jarlskog parameter could be expressed as
\begin{eqnarray}
J^L&=&2|(U_{\tau1})_{\rm KM}(U^*_{\tau3})_{\rm KM}||(U_{e1})_{\rm
KM}(U^*_{e3})_{\rm KM}|\sin\varphi_2\nonumber\\
&=&2|(U_{e1})_{\rm
KM}(U^*_{\tau1})_{\rm KM}||(U_{e3})_{\rm KM}(U^*_{\tau3})_{\rm
KM}|\sin\varphi'_2,\nonumber
\end{eqnarray}
with $\varphi_2=\varphi'_2$ as the common angle for the unitarity
boomerang, as shown in Fig.~\ref{fig3}.
\newpage

Fig.~\ref{fig3} is drawn under the approximation of tri-bimaximal matrix and $|U_{e3}|\sim0.1$ as an illustration so that $XY\sim YZ$, $XY'\sim Y'Z'$, and $XY\sim0.87XY'\sim3.5XZ\sim4XZ'$.

In the KM parametrization of the PMNS matrix,
\begin{eqnarray}
\varphi_2=\arg\left(-{s^L_1s^L_2(c_1^Ls^L_2s^L_3-c^L_2c^L_3e^{-i\delta^L_{\rm KM}})\over c^L_1(-s^L_1s^L_3)}\right).\nonumber
\end{eqnarray}

If $|U_{e3}|$ is small enough, $s^L_3\ll1$, we have $c_1^Ls^L_2s^L_3\ll1$, then
\begin{eqnarray}
\varphi_2 \approx  \arg\left({s^L_1 s^L_2(-c^L_2c^L_3e^{-i\delta^L_{\rm KM}})\over c^L_1s^L_1s^L_3 }\right) = \pi-\delta^L_{\rm KM}.\nonumber
\end{eqnarray}

If $|U_{e3}|$ is not so small, there should be
$\varphi_2<\pi-\delta^L_{\rm KM}$. The CP-violating phase
$\delta^L_{\rm KM}$ satisfies $\delta^L_{\rm KM}\lesssim\pi -
\varphi_2$ in the KM parametrization. So $\varphi_2\gtrsim74^\circ$ and implies that $\delta^L_{\rm KM}\lesssim106^\circ$ approximately.

\begin{figure}[hb]
\begin{center}
\scalebox{0.7}{\includegraphics[101,560][473,770]{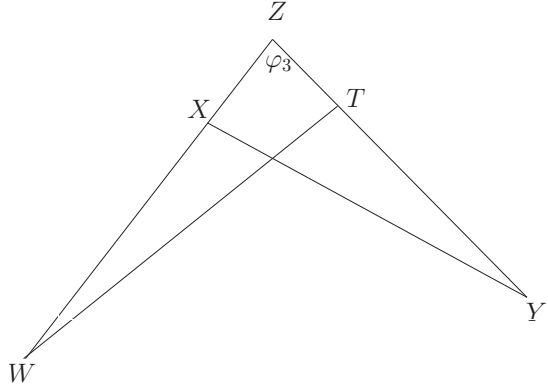}}
\caption{The unitarity boomerang in the lepton sector with the common
angle $\varphi_3$. The sides are: $XZ=|(U_{e1})_{\rm
CK}(U^*_{e3})_{\rm CK}|$, $XY=|(U_{\tau1})_{\rm
CK}(U^*_{\tau3})_{\rm CK}|$, $YZ=|(U_{\mu1})_{\rm
CK}(U^*_{\mu3})_{\rm CK}|$, $ZT=|(U_{e3})_{\rm CK}(U^*_{\mu3})_{\rm
CK}|$, $ZW=|(U_{e1})_{\rm CK}(U^*_{\mu1})_{\rm CK}|$ and
$TW=|(U_{e2})_{\rm CK}(U^*_{\mu2})_{\rm CK}|$. \label{fig4}}
\end{center}
\end{figure}

{\it Case 2'}. The unitarity boomerang in the CK parametrization. In
this case, we should introduce another unitarity triangle arising
from
\begin{eqnarray}
U_{e1}U^*_{\mu1}+U_{e2}U^*_{\mu2}+U_{e3}U^*_{\mu3}=0,\label{pmns3}
\end{eqnarray}
to manifest the CP-violating phase $\delta^L_{\rm CK}$ with the CK
parametrization. We present the Jarlskog parameter as
\begin{eqnarray}
J^L&=&2|(U_{e1})_{\rm CK}(U^*_{e3})_{\rm CK}||(U_{\mu1})_{\rm CK}(U^*_{\mu3})_{\rm CK}|\sin\varphi_3\nonumber\\
&=&2|(U_{e3})_{\rm CK}(U^*_{\mu3})_{\rm CK}||(U_{e1})_{\rm CK}(U^*_{\mu1})_{\rm CK}|\sin\varphi'_3,\nonumber
\end{eqnarray}
with $\varphi_3=\varphi'_3$ as the common angle for the unitarity
boomerang, as shown in Fig.~\ref{fig4}.

Here the unitarity triangle $XYZ$ is still the same one in
Fig.~\ref{fig3}. We draw Fig.~\ref{fig4} under the approximation of tri-bimaximal matrix and $|U_{e3}|\sim0.1$ for illustration, then $XY\sim YZ$, $ZW\sim TW$, and $XY\sim0.87ZW\sim3.5XZ\sim4ZT$.

The common angle $\varphi_3$ for the boomerang in the CK
parametrization is
\begin{eqnarray}
\varphi_3 = \arg\left(-{c^L_{12}c^L_{13}s^L_{13}e^{i\delta^L_{\rm CK}}\over (-s^L_{12}c^L_{23}-c^L_{12}s^L_{23}s^L_{13}e^{i\delta^L_{\rm CK}})s^L_{23}c^L_{13}}\right)\nonumber
\end{eqnarray}

If $|U_{e3}|$ is small enough, $s^L_{13}\ll1$, then $c^L_{12}s^L_{23}s^L_{13}\ll1$, we have
\begin{eqnarray}
\varphi_3\approx \arg\left({c^L_{12}c^L_{13}s^L_{13}e^{i\delta^L_{\rm CK}}\over s^L_{12}c^L_{23}s^L_{23}c^L_{13}}\right)=\delta^L_{\rm CK}.\nonumber
\end{eqnarray}

If $|U_{e3}|$ is not so small, we have
\begin{eqnarray}
\varphi_3=\arg(s^L_{12}c^L_{23}e^{i\delta^L_{\rm CK}}+c^L_{12}s^L_{23}s^L_{13})<\delta^L_{\rm CK}.\nonumber
\end{eqnarray}
Then we find $\delta^L_{\rm CK}\gtrsim\varphi_3$ in the CK parametrization. As an approximation, $\varphi_3\gtrsim74^\circ$ implies that $\delta^L_{\rm CK}\gtrsim74^\circ$.

We find that the two unitarity boomerangs are more or less of
similar sizes of their sides in the lepton sector in the special case, i.e. the tri-bimaximal matrix with $|U_{e3}|\lesssim0.2$ as an approximation of the PMNS matrix.
That is because all the elements in the PMNS matrix are of the same order of
magnitude except $U_{e3}$. In this special case, we can estimate the inner
angles of the unitarity triangles and give a constraint to the
CP-violating phase approximately. However, we can not get the isosceles triangles in Fig.~\ref{fig3} or Fig.~\ref{fig4} in common cases since the PMNS
matrix is not the tri-bimaximal matrix and the exact value of
$|U_{e3}|$ is unknown. When $|U_{e3}|$ is not small enough, the unitarity triangles in the unitarity boomerangs of Fig.~\ref{fig3} or Fig.~\ref{fig4} could still be nearly isosceles. But when $|U_{e3}|$ is small enough, any deviation between the two longer sides will cause a violation of the isosceles triangle. Then the inner angles of the unitarity triangles can not be determined and $\varphi_2$, $\varphi_3$, $\varphi'_2$, $\varphi'_3$ can cover the range of 0 to $\pi$. Nevertheless, $\delta^L_{\rm KM}\lesssim\pi -
\varphi_2$ in {\it Case 1'} and $\delta^L_{\rm CK}\gtrsim\varphi_3$ in {\it Case 2'} still hold although there is no numerical constraint on the CP violating phases in both cases.

When discussing unitarity triangles or boomerangs in the lepton sector, we do not take the phase matrix $P={\rm
diag}(e^{i\alpha_1/2},e^{i\alpha_2/2},1)$ into account. The reason is that the phase matrix $P$ does not affect the unitarity of the PMNS matrix and $V_{\rm PMNS}
V^\dagger_{\rm PMNS} = V^\dagger_{\rm PMNS} V_{\rm PMNS} = I$ (where $I$ is the $3\times3$ unitary matrix) though the neutrinos are of Majorana type. For instance, the unitarity triangle $XYZ$ arises from $U_{e1}U^*_{e3}+U_{\mu1}U^*_{\mu3}+U_{\tau1}U_{\tau3}^*=0$. If neutrinos are of Majorana type, the unitarity triangle $XYZ$ will arise from $(U_{e1}U^*_{e3}+U_{\mu1}U^*_{\mu3}+U_{\tau1}U_{\tau3}^*)e^{i\alpha_1/2}=0$. However, the phase does not take effect here and we can still consider the unitarity triangles or boomerangs with Dirac neutrinos for the PMNS matrix.

In the lepton sector, though there is not any evident difference
among the six unitarity triangles since the elements of the PMNS
matrix are of the same order of magnitude, we prefer to choose the
three unitarity triangles arising from Eq.~(\ref{pmnstwo}) and
Eq.~(\ref{pmns3}) corresponding to the quark sector because they are
convenient for analyzing and comparing, especially when we wish to
analyse possible relations between quarks and leptons, such as the
quark-lepton complementarity discussed in Sec.~II.

\section{Summary}

In this work, we have studied the mixing matrices of quarks and
leptons for two cases of KM parametrization and CK parametrization,
which express the mixing matrices with three mixing angles and one
CP-violating phase. We present the transformations between the two
cases of parametrizations and obtain the relations between their
parameters. We also find that the quark-lepton complementarity (QLC)
revealed in the CK parametrization is still well kept in the KM
parametrization when $s^L_3\sim \mathcal {O}(10^{-2})$ or smaller,
i.e. the value of $|U_{e3}|$ is small enough. Then we analyse the
unitarity boomerangs, a new concept
proposed by Frampton and He for convenient study of the quark mixing
matrix, under both the KM and CK parametrizations in the quark
sector and extend the idea of unitarity boomerang to the lepton
sector. With help of the unitarity boomerang, we analysed the
constraints of the Dirac CP-violating phase in KM and CK
parametrizations in both quark and lepton sectors.
Our study is helpful for a comprehensive
understanding of the mixing patterns and properties for both quarks
and leptons from a unified viewpoint.

\section*{Acknowledgments}
This work is partially supported by National
Natural Science Foundation of China (No. 10721063 and No. 10975003),
by the Key Grant Project of Chinese Ministry of Education (No.
305001), and by the Research Fund for the Doctoral Program of Higher
Education (China).

\end{document}